\documentclass[aps,pra,reprint,
twocolumn,longbibliography]{revtex4-1}
\usepackage{graphicx}
\usepackage{bm}
\usepackage{hyperref}
\usepackage{color}
\usepackage{dcolumn}
\usepackage[utf8]{inputenc} 
\usepackage{amsmath}
\usepackage{braket}
\usepackage{soul} 
\usepackage{xcolor}
\usepackage{epstopdf}

\begin{document}


\title{Enhancing non-classical correlations for light scattered by an ensemble  of cold two-level atoms}

\author{Lucas S. Marinho}
\affiliation{Departamento de Física, Universidade Federal de Pernambuco, 50670-901 Recife, Pernambuco, Brazil}

\author{Michelle O. Ara\'ujo}
\affiliation{Departamento de Física, Universidade Federal de Pernambuco, 50670-901 Recife, Pernambuco, Brazil}

\author{Wellington Martins}
\affiliation{Departamento de Física, Universidade Federal de Pernambuco, 50670-901 Recife, Pernambuco, Brazil}

\author{Daniel Felinto}
\email{daniel.felinto@ufpe.br}
\affiliation{Departamento de Física, Universidade Federal de Pernambuco, 50670-901 Recife, Pernambuco, Brazil}

\begin{abstract}
We report the enhancement of quantum correlations for
biphotons generated via spontaneous four-wave mixing in
an ensemble of cold two-level atoms. This enhancement
is based on the filtering of the Rayleigh linear component
of the spectrum of the two emitted photons, favoring
the quantum-correlated sidebands reaching the detectors.
We provide direct measurements of the unfiltered spectrum
presenting its usual triplet structure, with Rayleigh
central components accompanied by two peaks symmetrically
located at the detuning of the excitation laser with
respect to the atomic resonance. The filtering of the central
component results in a violation of the Cauchy–Schwarz
inequality to $(4.8 \pm 1.0) \leq 1$ for a detuning of 60 times the
atomic linewidth, representing an enhancement by a factor
of four compared with the unfiltered quantum correlations
observed at the same conditions.

\vspace{0.5cm}

\textit{Received 27 April 2023; revised 18 May 2023; accepted 20 May 2023; posted 23 May 2023; published 13 June 2023}
\end{abstract}


\maketitle

The nonlinear phenomenon of four-wave mixing (FWM), in which the interactions between three optical fields inside a medium generate a new fourth field, has many applications in optical phase conjugation~\cite{He2002}, parametric amplification~\cite{Khudus2016}, supercontinuum generation~\cite{Dudley2006}, optical frequency combs~\cite{DelHaye2007}, among others. It has been also a quite successful source for optical fields presenting quantum correlations~\cite{Slusher1985,Maeda1987,Raizen1987, Lambrecht1996,McCormick2007}. After the proposal of the Duan-Lukin-Cirac-Zoller (DLCZ) protocol for long-distance quantum communication~\cite{Duan2001}, spontaneous FWM (SFWM) has been particularly explored by many groups as a source of quantum-entangled biphotons from ensembles of multi-level atoms \cite{Kuzmich2003,Balic2005,Matsukevich2005, Thompson2006,Yuan2008,Albrecht2015,Ortiz2018}.

A challenging aspect of SFWM is that the quantum correlated fields are weak when compared with linear scattering by the excitation fields. This linear scattering corresponds to the elastic Rayleigh component of the overall scattered light. Typically, it is then required to filter out this lowest-order component in order to access the quantum correlations. This filtering is done through the use of a different frequency or polarization for the biphoton with respect to the excitation light, as in Refs.~\cite{Kuzmich2003,Balic2005,Matsukevich2005, Thompson2006,Yuan2008,Albrecht2015, Ortiz2018}, implying atomic excitations through a level structure of at least three states. In 2007, however, Du \textit{et al.} theoretically predicted the possibility of observing quantum correlations in SFWM from an ensemble of two-level atoms even without the use of any filter~\cite{Du2007}. In this case, coincidences coming from quantum correlations would prevail, directly, over coincidences coming from two independent Rayleigh scattering processes. In 2022, this prediction was experimentally demonstrated~\cite{Araujo2022}, highlighting the strength of quantum correlations in simpler, ubiquitous systems.

The nonlinear interaction of two-level atoms with light has been well studied since the beginning of the field of nonlinear optics, as it represents the first approximation for multiple systems presenting strong interactions close to a resonance~\cite{Allen1987,Boyd2003}. The observation of nonclassical correlations in such systems allows for the application of known schemes to address higher order nonlinearities in ensembles of two-level atoms~\cite{Raj1984}, aiming to implement new classes of light fields with nonclassical correlations. However, the degree of quantum  correlations reported in Ref.~\cite{Araujo2022}, even though consistent with the degree predicted in theory for unfiltered light, is quite small. This fact effectively prevents the exploration of the effect in other directions, particularly for generating higher order nonlinear processes.

In the present work, thus, we implement a strategy to enhance the quantum correlations in a system. First, we measure the spectrum of the fields generated by unfiltered SFWM and observe the appearance of three frequency peaks, with photons generated at the central component (Rayleigh scattering) and at the sidebands shifted by the generalized Rabi frequency to the atomic resonance. This spectrum is mostly determined by the third-order nonlinear susceptibility $\chi^{(3)}$ of the medium, as previously discussed in connection with the theoretical treatment for the process~\cite{Wen2007}. Then, introducing two frequency filters to reduce the contribution from the Rayleigh scattering~\cite{Aspect1980}, while still preserving the bandwidth required to observe short-lived correlations, we finally show the improvement of quantum correlations as the Rayleigh component is increasingly suppressed. This advance establishes a clear pathway to apply biphotons generated from two-level systems in multiple directions. Besides the exploration of the aforementioned higher order nonlinearities, we are now able to employ the strong cycling transitions for efficient generation of narrowband biphotons, aiming applications, for example, in quantum communication.

\begin{figure}[t]
\centering
\includegraphics[width=8cm]{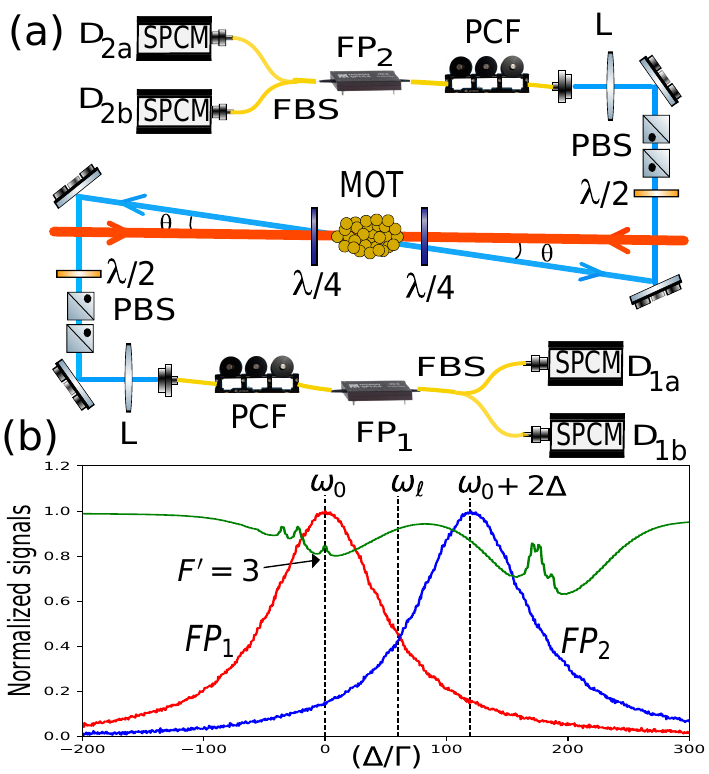}
\caption{(a) Experimental setup for spontaneous four-wave mixing in a cold ensemble of pure two-level atoms. (b) Fabry-Perot filters transmission signals (red and blue lines) as a function of detuning, using a spectrum of saturated absorption as reference (green line) with the $F'=3$ peak indicating the resonance for the transition  $5S_{1/2}(F=2) \rightarrow 5P_{3/2}(F'=3)$ of Rb$^{87}$. MOT, magneto-optical trap; PCF, polarisation controller fiber; PBS, polarising beam splitter; SPCM, single-photon counting model; FBS, fiber beam splitter; L, lens; FP, Fabry-Perot;
$\lambda /2$ ($\lambda /4$), half-wave (quarter-wave) plate. }
\label{fig1}
\end{figure}

The experimental scheme for the generation of SFWF is shown in
Fig. \ref{fig1}(a), with two counter-propagating pump fields (in orange) of same power $P$
spontaneously generating pairs of photons (fields 1 and 2, in
blue) emitted in opposite directions, forming an angle $\theta = 3^{\circ}$
with the pump fields.  We prepare an ensemble of cold $^{87}$Rb atoms from a magneto-optical trap, with optical depth around $OD \approx 15$. After turning off the trap laser and magnetic field, the repumper laser is kept on for an extra 900 $\mu$s to allow for the preparation of all atoms in the $5S_{1/2}(F=2)$ hyperfine ground state. Then 50 $\mu$s after the repumper is turned off, the pump fields are turned on during 1 ms, with the whole trap operating cycle repeated at every 25 ms.

The two pump fields have the same frequency $\omega_l=\omega_0+\Delta$, with $\Delta$ the detuning from the frequency $\omega_0$ of the  $5S_{1/2}(F=2) \rightarrow 5P_{3/2}(F'=3)$ transition. They have the same circular $\sigma^+$  polarisation as seen by the atoms, resulting in optical pumping to the $5S_{1/2}(F=2,m_F=2)$ Zeeman sublevel  and the subsequent excitation of just the pure two level transition $5S_{1/2}(F=2,m_F=2) \rightarrow 5P_{3/2}(F'=3,m_{F'}=3)$~\cite{Araujo2022}. The photons are collected using single mode fibers. After emission, they go through $\lambda/4$ plates transforming their polarization from circular to linear. The photons' degree of linear polarization is then verified to be $(99.0\pm 0.2)\%$. The $4\sigma$ diameters for the spatial modes at the ensemble are 420 $\mu$m and 140 $\mu$m for pump fields and detection modes, respectively.

In this backward SFWM process, as demonstrated in Refs. \cite{Du2007, Wen2007}, the emitted photons have frequencies $\omega_1$ and $\omega_2$, equal to $\omega_l$ (Rayleigh component) or $\omega_l \pm \Delta$ (sidebands). The interference between these spectral components is responsible for the observation of oscillations in the intensity correlations functions $g_{ij}(t,t+\tau)=\langle I_{D_i}(t)I_{D_j}(t+\tau) \rangle / \langle I_{D_i}(t)\rangle \langle I_{D_j}(t+\tau) \rangle $ between fields $i$ and $j$, with $I_{D_i}(t)$ the intensity of light measured at a detector $D_i$ for field $i$ ($i=1a,1b,2a,2b$) at time $t$ counted from the moment the pump fields are turned on. $\langle \dots \rangle$ denotes an ensemble average over many samples of pump periods. $\tau$ is the time delay for detections in $D_i$ and $D_j$. The detectors are fiber-coupled single-photon counting modules (model SPCM-AQRH-13-FC from Perkin Elmer) with outputs directed to a multiple-event time digitiser with 100 ps time resolution (model MCS6A from FAST ComTec).  

\begin{figure}[t]
\centering
\includegraphics[width=\linewidth]{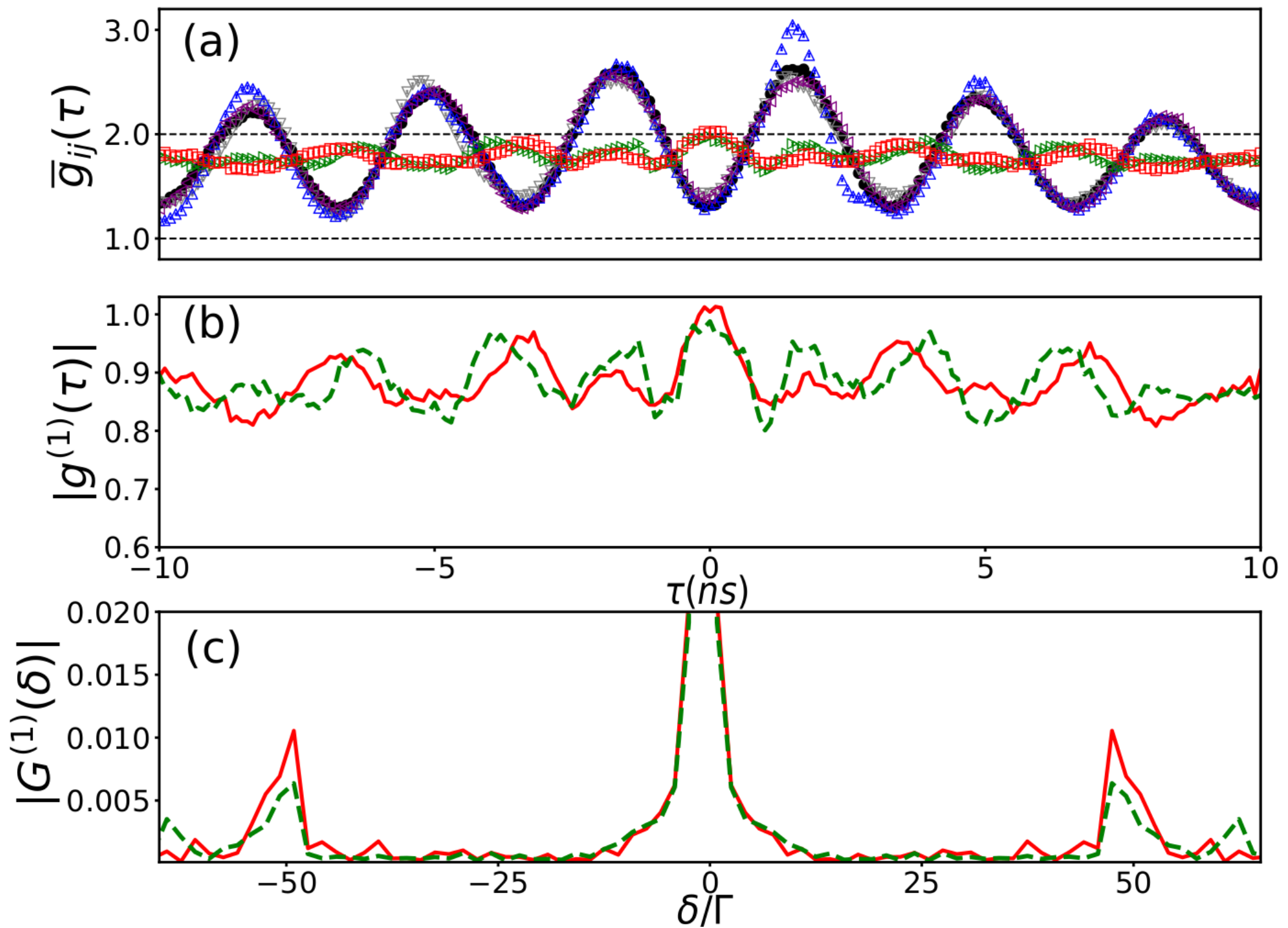}
\caption{(a) Second order correlations functions $\overline{g}_{ij}^{(2)}(\tau)$, with: 
$\overline{g}_{2a1a}$ (gray), $\overline{g}_{2b1a}$ (black), $\overline{g}_{1b2a}$ (blue), $\overline{g}_{2b1b}$ (violet), $\overline{g}_{1b1a}$ (green) and $\overline{g}_{2b2a}$ (red). (b) Temporal dependence of first-order correlation functions $|g^{(1)}(\tau)|$ for fields 1 (dashed green curve) and 2 (continuous red curve),  and (c) their correspondent fast Fourier transforms normalised by their maximum values. Other parameters are $P=350$ $\mu$W and $\Delta=50\Gamma$.}
\label{fig2}
\end{figure}

Figure \ref{fig2}(a) shows these oscillations in the temporal dependence of all the second-order correlation functions, in a short timescale of up to 10$\,$ns. To improve the data statistics,  we average the correlation functions during the whole $T=1$ ms trial period, defining $\overline{g}_{ij}(\tau)=\frac{1}{T}\int_{-T/2}^{T/2}g_{ij}(t,t+\tau)dt$~\cite{Araujo2022}. The correlation functions between photons from different fields are called cross-correlations, represented by $\overline{g}_{2a1a}$ (gray), $\overline{g}_{2b1a}$ (black), $\overline{g}_{1b2a}$ (blue) and $\overline{g}_{2b1b}$ (violet). On the other hand, the correlations between photons from the same field are called auto-correlations, represented by $\overline{g}_{1b1a}$ (green) and $\overline{g}_{2b2a}$ (red). All these second-order correlation functions gradually decrease to unity, indicating lack of correlations, in a long timescale on the order of 40$\,\mu$s~\cite{Moreira2021, Marinho2023}.

The photon spectrum is obtained from the auto-correlation functions $\overline{g}_{1b1a}(\tau)$ and  $\overline{g}_{2b2a}(\tau)$ using the Siegert relation \cite{Loudon1983, Eloy2018}, valid for fields with thermal statistics:
 
\begin{equation}
    g^{(2)}(\tau) = 1 + |g^{(1)}(\tau)|^2 \,,
\end{equation}
\\*
which relates the second-order auto-correlation function $g^{(2)}(\tau)$ with the first order correlation function $g^{(1)}(\tau)$. Figure \ref{fig2}(b) plots then the first order correlation functions for fields 1 and 2, showing that they still preserve the oscillatory behavior of the respective auto-correlation functions. From a fast Fourier transform of Fig.~\ref{fig2}(b), Fig. \ref{fig2}(c) finally shows the spectra of fields 1 and 2. These spectra follow the predictions of Ref. \cite{Wen2007}, with photons mainly generated at the central component or at the two sidebands. Note that the sidebands are two orders of magnitude smaller than the central component.

\begin{figure}[t]
\centering
\includegraphics[width=\linewidth]{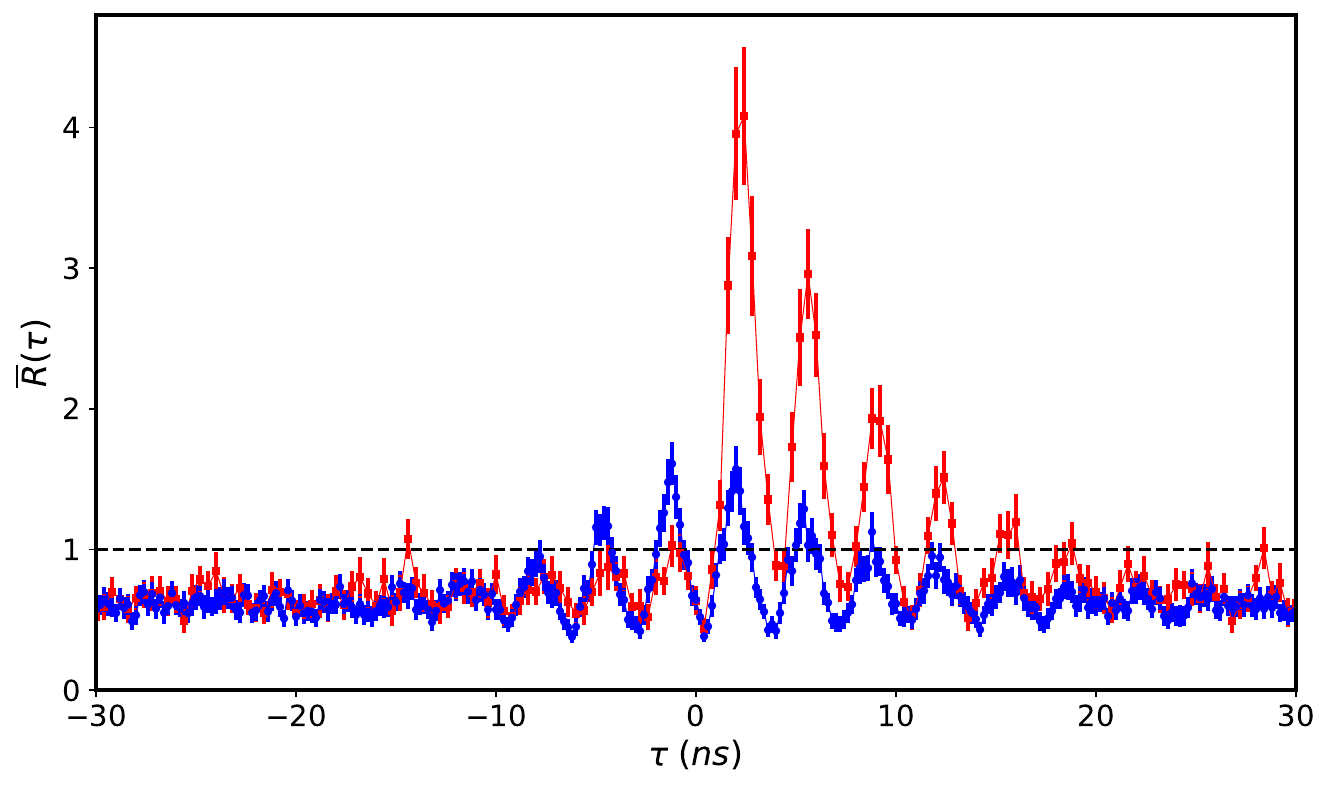}
\caption{Behavior of the degree of correlation $\overline{R}$ as a function of $\tau$ on the conditions of Fig.~\ref{fig2}. The blue-filled (red-filled) circles (squares) are the results without (with) Fabry-Perot filters. The lines connecting the points are just guides for the eyes. The black dashed line represents the classical limit of the Cauchy-Schwarz inequality.}
\label{fig3}
\end{figure}

In order to reduce the contribution of Rayleigh scattering to the correlation functions, we then used two Fabry-Perot filters in fibers (model FFP-I from Micron Optics) operating at 780 nm, with full width at half maximum FWHM$\approx$ 600 MHz, free spectral range FSR $\approx 20$ GHz, finesse $ \mathcal{F}=$ FSR/FWHM = 33, and typical insertion loss of around 3 dB. Since the photons from the sidebands have frequencies $\omega_{1,2}=\omega_l \pm \Delta$, corresponding to $\omega_0$ (resonance) or $\omega_0 +2\Delta$ (out off-resonance), the filters were positioned at these frequencies, as can be seen in Fig. \ref{fig1}(b). The r$  $ed and blue Fabry-Perot filter transmission signals were calibrated in frequency using the saturated absorption spectrum, with the $F'=3$ peak denoting the resonance for the transition $5S_{1/2}(F=2) \rightarrow 5P_{3/2}(F'=3)$. The relatively large bandwidth of the filters allows for the observation of the fast quantum correlations in the system, on the scale of nanoseconds. 

Our criterion for classifying quantum correlations is the degree of violation of the Cauchy-Schwarz inequality $R \leq 1$ valid for classical fields~\cite{Clauser1974}, in which $R$, in our setup, depends on $\tau$ and has two different expressions corresponding to equivalent combinations of the second-order correlation functions: 

\begin{equation}\label{R1_R2}
R_1(\tau)= \frac{\overline{g}_{2b1a}(\tau) \overline{g}_{2a1b}(\tau)}{\overline{g}_{1b1a}(0) \overline{g}_{2b2a}(0)}, R_2(\tau)= \frac{\overline{g}_{2a1a}(\tau) \overline{g}_{2b1b}(\tau)}{\overline{g}_{1b1a}(0) \overline{g}_{2b2a}(0)}.
\end{equation}

In the following, for clarity we will focus on the average $\overline{R}(\tau)$ of these two quantities. Figure \ref{fig3} shows $\overline{R}(\tau)$ as a function of $\tau$ for the parameters of Fig.~\ref{fig2}, with the black dashed line representing the maximum correlations R = 1 that can be reached classically. The red (blue) points are the values of $\overline{R}(\tau)$ with (without) the Fabry-Perot filters. The maximum degree of correlation  $\overline{R}_{max}$ is then significantly enhanced once we include the filters to attenuate the central component of the spectrum. For the red points, the Fabry-Perot 1 ($\mathrm{FP}_1$) was positioned at resonance  $\omega_0$ and the Fabry-Perot 2 ($\mathrm{FP}_2$) was out of resonance, at $\omega_0 +2\Delta$. Notice that the quantum correlations appear only for values of $\tau>0$. We also checked that on inverting the tuning of the filters, i.e., putting $\mathrm{FP}_1$ out of resonance and $\mathrm{FP}_2$ at resonance, the quantum correlation appears only for values $\tau<0$. Therefore, the photons out of resonance are always emitted first, as they do not suffer any delays coming from absorptions and re-emissions by the atoms~\cite{Aspect1980}. The differences between the two curves in Fig.~\ref{fig3} come from modifications of the cross-correlation functions. The auto-correlation functions were not significantly affected by the inclusion of the filters, as they still behave as expected for thermal states.

\begin{figure}[t]
\centering
\includegraphics[width=\linewidth]{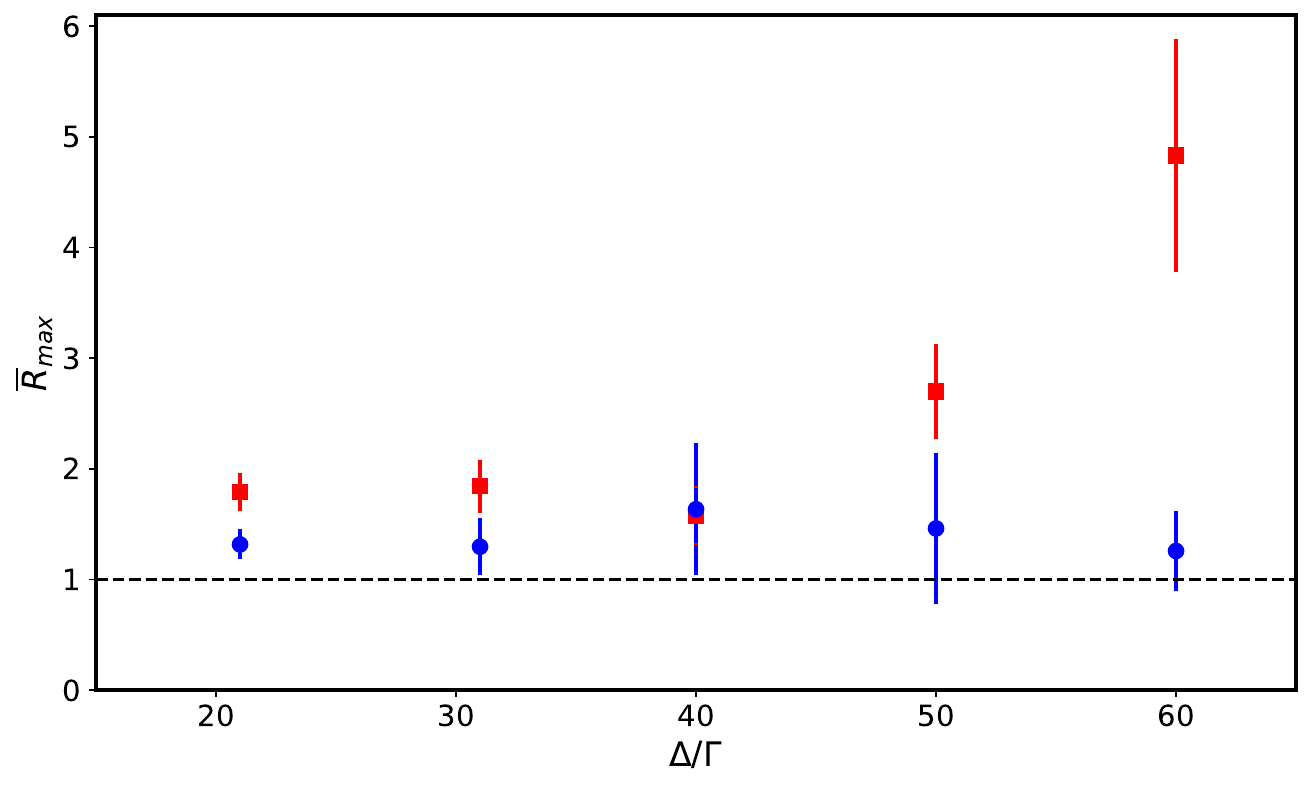}
\caption{Maximum degree of correlation $\overline{R}_{max}$ as a function of the detuning $\Delta$. Blue-filled circles (red-filled squares) correspond to measures without (with) the use of Fabry-Perot filters. The pump power was fixed at $P=330$ $\mu$W.}
\label{fig4}
\end{figure}

The collection efficiency without filters is estimated to be around 40$\%$, the product of the fiber's coupling efficiency (70$\%$) and the detector's efficiency (60$\%$). It goes down to 20$\%$ with filters, as our Fabry-Perot filters have an insertion loss of around 3$\,$dB. For the conditions of Fig.~\ref{fig3}, the coincidence rate is of around 300$\,$Hz without filters, considering a delay window of 20$\,$ns. This value goes down to 13$\,$Hz with filters. Even though this overall coincidence rate is low so far, it has much room for improvement, as narrower filters with better insertion loss should allow operation closer to resonance with much higher rates. This value should also be improved with higher optical depths and pump powers, as we operated here at the maximum values of these two quantities for our system. Also, for comparison with other sources, we must note the narrowband (tens of megahertz) nature of the generated biphotons and its pulsed operation (1 out of 25$\,$ms).

Figure \ref{fig4} shows $\overline{R}_{max}$ for different values of $\Delta$ with and without the use of Fabry-Perot filters. The curve in Fig. \ref{fig4} is not optimized (in terms of alignment) for the correlations, as can be seen from a comparison of similar conditions in Fig. \ref{fig3} and Ref.~\cite{Araujo2022}, but it provides a continuous series of measurements with different detunings. The increasing detunings imply decreasing levels $\alpha$ of the Rayleigh component in the detected modes, with this level going from $\alpha =  0.87$ (for $\Delta = 20\Gamma$) to $\alpha = 0.45$ (for $\Delta = 50\Gamma)$, with $\alpha$ given by the transmission at $\omega_l$ for the Fabry-Perot curves in Fig.~\ref{fig1}(b). On the other hand, the larger detunings imply smaller count rates, for our available maximum pump power, and more sensitivity of the system to fluctuations during longer periods of data collection. Even though, we were still able to obtain an enhancement factor with filters of the order of 4 for $\overline{R}_{max}$ when compared with the result without filters under the same conditions, reaching $\overline{R}_{max} = 4.8 \pm 1.0$ for $\Delta = 60\Gamma$.

In conclusion, we measured the spectra of the individual photons in the process of biphoton generation from an ensemble of pure two-level atoms, and directly observed the sidebands carrying the core of the quantum correlations in the system. In order to enhance these correlations, we filtered the central, Rayleigh component of the spectra. After this procedure, we observed an increase by a factor of four in the Cauchy-Schwarz inequality violation for the system. We also verified the time ordering of the emission, with the out-of-resonance photon always leaving the ensemble first. These results provide a clear strategy for future improvements in the distillation of quantum correlations in the system, requiring a combination of larger detunings, filters with narrower bandwidths, and higher powers for the pump fields. This advance effectively enables the generation of quantum correlated biphotons from a large class of systems that can be approximated as pure two-level atoms.

\textbf{Funding.} Conselho Nacional de Desenvolvimento Científico e Tecnológico
(465469/2014-0); Coordenação de Aperfeiçoamento de Pessoal de Nível Superior (23038.003069/2022-87); Fundação de Amparo à Ciência e Tecnologia do Estado de Pernambuco; Fundação de Amparo à Pesquisa do Estado de São Paulo (2021/06535-0); Office of Naval Research
(N62909-23-1-2014).

\textbf{Disclosures.} The authors declare no conflicts of interest.

\textbf{Data availability.} Data underlying the results presented in this paper are
not publicly available at this time but may be obtained from the authors upon
reasonable request.



\end{document}